\documentclass[aps,prd,twocolumn,showpacs,preprintnumbers,superscriptaddress,tightenlines,amsmath,amssymb]{revtex4}

\usepackage{graphicx} 
\usepackage{dcolumn}  
\usepackage{color}
\usepackage{ulem}
\usepackage{pstricks}

\graphicspath{{ps}}

\renewcommand{\thefootnote}{\fnsymbol{footnote}}

\begin{document}

\title{ \quad\\[0.5cm]  Measurements of time-dependent $CP$ violation in $B^0 
\to \psi(2S) K_S^0$ decays}

\affiliation{Budker Institute of Nuclear Physics, Novosibirsk}
\affiliation{University of Cincinnati, Cincinnati, Ohio 45221}
\affiliation{Department of Physics, Fu Jen Catholic University, Taipei}
\affiliation{The Graduate University for Advanced Studies, Hayama}
\affiliation{Hanyang University, Seoul}
\affiliation{University of Hawaii, Honolulu, Hawaii 96822}
\affiliation{High Energy Accelerator Research Organization (KEK), Tsukuba}
\affiliation{Institute of High Energy Physics, Chinese Academy of Sciences, Beijing}
\affiliation{Institute of High Energy Physics, Vienna}
\affiliation{Institute of High Energy Physics, Protvino}
\affiliation{Institute for Theoretical and Experimental Physics, Moscow}
\affiliation{J. Stefan Institute, Ljubljana}
\affiliation{Kanagawa University, Yokohama}
\affiliation{Korea University, Seoul}
\affiliation{Kyungpook National University, Taegu}
\affiliation{\'Ecole Polytechnique F\'ed\'erale de Lausanne (EPFL), Lausanne}
\affiliation{Faculty of Mathematics and Physics, University of Ljubljana, Ljubljana}
\affiliation{University of Maribor, Maribor}
\affiliation{University of Melbourne, School of Physics, Victoria 3010}
\affiliation{Nagoya University, Nagoya}
\affiliation{Nara Women's University, Nara}
\affiliation{National Central University, Chung-li}
\affiliation{National United University, Miao Li}
\affiliation{Department of Physics, National Taiwan University, Taipei}
\affiliation{H. Niewodniczanski Institute of Nuclear Physics, Krakow}
\affiliation{Nippon Dental University, Niigata}
\affiliation{Niigata University, Niigata}
\affiliation{University of Nova Gorica, Nova Gorica}
\affiliation{Osaka City University, Osaka}
\affiliation{Osaka University, Osaka}
\affiliation{Saga University, Saga}
\affiliation{University of Science and Technology of China, Hefei}
\affiliation{Seoul National University, Seoul}
\affiliation{Sungkyunkwan University, Suwon}
\affiliation{University of Sydney, Sydney, New South Wales}
\affiliation{Tata Institute of Fundamental Research, Mumbai}
\affiliation{Toho University, Funabashi}
\affiliation{Tohoku Gakuin University, Tagajo}
\affiliation{Tohoku University, Sendai}
\affiliation{Department of Physics, University of Tokyo, Tokyo}
\affiliation{Tokyo Institute of Technology, Tokyo}
\affiliation{Tokyo Metropolitan University, Tokyo}
\affiliation{Tokyo University of Agriculture and Technology, Tokyo}
\affiliation{Virginia Polytechnic Institute and State University, Blacksburg, Virginia 24061}
\affiliation{Yonsei University, Seoul}
  \author{H.~Sahoo}\affiliation{University of Hawaii, Honolulu, Hawaii 96822} 
  \author{T.~E.~Browder}\affiliation{University of Hawaii, Honolulu, Hawaii 96822} 
  \author{K.~Trabelsi}\affiliation{High Energy Accelerator Research Organization (KEK), Tsukuba} 
  \author{I.~Adachi}\affiliation{High Energy Accelerator Research Organization (KEK), Tsukuba} 
  \author{H.~Aihara}\affiliation{Department of Physics, University of Tokyo, Tokyo} 
  \author{K.~Arinstein}\affiliation{Budker Institute of Nuclear Physics, Novosibirsk} 
  \author{T.~Aushev}\affiliation{\'Ecole Polytechnique F\'ed\'erale de Lausanne (EPFL), Lausanne}\affiliation{Institute for Theoretical and Experimental Physics, Moscow} 
  \author{S.~Bahinipati}\affiliation{University of Cincinnati, Cincinnati, Ohio 45221} 
  \author{A.~M.~Bakich}\affiliation{University of Sydney, Sydney, New South Wales} 
  \author{V.~Balagura}\affiliation{Institute for Theoretical and Experimental Physics, Moscow} 
% ...

\collaboration{Belle Collaboration}

\noaffiliation

\begin{abstract}
We report improved measurements of time-dependent $CP$ violation
parameters for $B^0 (\overline{B}{}^0) \to \psi(2S) K_S^0$.
This analysis is based on a data sample of $657 \times 10^6$ $B\overline{B}$ 
pairs collected at the $\Upsilon(4S)$ resonance with the Belle detector 
at the KEKB energy-asymmetric $e^+ e^-$ collider. 
We fully reconstruct
one neutral $B$ meson in the $\psi(2S) K_S^0$ $CP$-eigenstate 
decay channel, 
and the flavor of the accompanying $B$ meson is identified to be 
either $B^0$ or $\overline{B}{}^0$ from its decay products.
$CP$ violation parameters are obtained from the asymmetries 
in the distributions of
the proper-time intervals between the two $B$ decays:
${\cal S}_{\psi(2S)K_S^0} = +0.72 \pm 0.09 (\rm stat) \pm 0.03 (\rm syst)$,
${\cal A}_{\psi(2S)K_S^0} = +0.04 \pm 0.07 (\rm stat) \pm 0.05 (\rm syst)$.
These results are in agreement with results from measurements of $B^0 \to J/\psi K^0$.
\end{abstract}

\pacs{11.30.Er, 12.15.Hh, 13.25.Hw}

\maketitle


{\renewcommand{\thefootnote}{\fnsymbol{footnote}}}
\setcounter{footnote}{0}
In the standard model, $CP$ violation in $B^0$
meson decays originates from an irreducible complex phase in the $3 \times 3$ 
Cabibbo-Kobayashi-Maskawa (CKM) mixing matrix~\cite{ckm}.
In the decay chain $\Upsilon(4S) \to B^0 \overline{B}{}^0 \to 
f_{CP}f_{\rm tag}$, where one of the $B$ mesons decays at 
time $t_{CP}$ to a $CP$ eigenstate $f_{CP}$ and the other decays
at time $t_{\rm tag}$ to a final state $f_{\rm tag}$ that distinguishes
between $B^0$ and $\overline{B}{}^0$, the decay rate has a time 
dependence~\cite{carter} given by
\begin{eqnarray}
\label{eq_decay}
{\cal P}(\Delta{t})= \frac{ e^{-|\Delta{t}|/{\tau_{B^0}}} }{4\tau_{B^0}}
\biggl\{1 & + & q \cdot 
 \Bigl[ {\cal S}_{f_{CP}} \sin(\Delta m_d \Delta{t})  \nonumber \\
 & + & {\cal A}_{f_{CP}} \cos(\Delta m_d \Delta{t})
\Bigr] \biggr\}.
\end{eqnarray}
Here ${\cal S}_{f_{CP}}$ and ${\cal A}_{f_{CP}}$ are the $CP$ 
violation parameters, 
$\tau_{B^0}$ is the neutral $B$ lifetime,
$\Delta m_d$ is the mass difference between the two 
neutral $B$ mass eigenstates, 
$\Delta t = t_{CP} - t_{\rm tag}$, and the $b$-flavor charge 
$q$ equals $+1$ $(-1)$ when the tagging $B$ meson is a 
$B^0$ ($\overline{B}{}^0$). 
For $f_{CP}$ final states resulting from a
$b \to c \overline{c} s$ transition, 
the standard model predicts ${\cal S}_{f_{CP}} = -\xi_f \sin 2 \phi_1$
\cite{beta} and 
${\cal A}_{f_{CP}} \simeq 0$, where $\xi_f = +1$ $(-1)$ for $CP$-even 
($CP$-odd) final states and $\phi_1$ is one of the three interior angles 
of the CKM unitarity triangle,
defined as $\phi_1 \equiv \pi - \arg(V_{tb}^*V_{td}/V_{cb}^*V_{cd})$.
Measurements of $CP$ asymmetries in $b \to c \overline{c} s$ 
transitions have been reported by Belle~\cite{belle_cc,belle_new} and 
BaBar~\cite{babar_cc}. Results from our previously published 
$B^0 \to \psi(2S) K_S^0$ analysis were based on a 140~fb$^{-1}$ data sample
corresponding to $152 \times 10^6$ $B \overline{B}$ pairs~\cite{belle_cc}. Here
we report new measurements 
with an improved analysis~\cite{impro}
incorporating an additional 465~fb$^{-1}$ 
data sample for a total of 605~fb$^{-1}$ ($657 \times 10^6$ $B \overline{B}$ 
pairs).
\par At the KEKB energy-asymmetric $e^+ e^-$ (3.5 on 8.0 GeV) 
collider~\cite{kekb}, the $\Upsilon(4S)$ is produced with a Lorentz 
boost of $\beta \gamma = 0.425$ nearly along the $z$ axis, which 
is defined as opposite to the positron beam direction.
Since the $B^0$ and $\overline{B}{}^0$ are approximately 
at rest in the $\Upsilon(4S)$ center-of-mass system (cms),
$\Delta t$ can be determined from the displacement in $z$ between the two 
decay vertices: $\Delta t \simeq \Delta z / (\beta \gamma c)$,
where $c$ is the speed of light.
%
\par The Belle detector is a large-solid-angle magnetic
spectrometer that consists of a silicon vertex detector (SVD),
a 50-layer central drift chamber (CDC), an array of
aerogel threshold 
Cherenkov counters (ACC),
a barrel-like arrangement of time-of-flight
scintillation counters (TOF), and an electromagnetic calorimeter (ECL)
comprised of CsI(Tl) crystals located inside
a superconducting solenoid coil that provides a 1.5~T
magnetic field.  An iron flux-return located outside 
the coil is instrumented to detect $K_L^0$ mesons and to identify
muons (KLM).  The detector is described in detail elsewhere~\cite{Belle}.
Two different inner detector configurations were used. For the first sample
of $152 \times 10^6$ $B\overline{B}$ pairs, a 2.0 cm radius beampipe
and a 3-layer silicon vertex detector (SVD-I) were used;
for the latter $505 \times 10^6$ $B\overline{B}$ pairs,
a 1.5 cm radius beampipe, a 4-layer silicon detector (SVD-II),
and a small-cell inner drift chamber were used~\cite{Natkaniec}.
%
%
\par We reconstruct $\psi(2S)$ mesons in the $l^+ l^-$ decay channel
($l = e$ or $\mu$) and $J/\psi \pi^+ \pi^-$ decay channel.
$J/\psi$ mesons are reconstructed in the $l^+ l^-$ decay channel
and include the bremsstrahlung photons that are within 50 mrad
of each of the $e^+$ and $e^-$ tracks [denoted as $e^+ e^- (\gamma)$].
The invariant mass of the $J/\psi$ candidates is required to be within 
$-0.150$ GeV/$c^2$ 
$< M_{e^+ e^- (\gamma)} - m_{J/\psi} < + 0.036$ GeV/$c^2$ and 
$-0.060$ GeV/$c^2 < M_{\mu^+ \mu^-} - m_{J/\psi} < + 0.036$ GeV/$c^2$,
where $m_{J/\psi}$ denotes the world-average $J/\psi$ mass~\cite{pdg2006}, and 
$M_{e^+ e^- (\gamma)}$ and $M_{\mu^+ \mu^-}$ are the reconstructed 
invariant masses of the $e^+ e^- (\gamma)$ and 
$\mu^+ \mu^-$ candidates, respectively. 
For the $\psi(2S) \to l^+ l^-$ candidates, 
the same procedure is used. In this case,
the invariant mass is required 
to be within $-0.150$ GeV/$c^2$ 
$< M_{e^+ e^- (\gamma)} - m_{\psi(2S)} < + 0.036$ 
GeV/$c^2$ and $-0.060$ GeV/$c^2 < M_{\mu^+ \mu^-} - m_{\psi(2S)} < + 0.036$ 
GeV/$c^2$, where $m_{\psi(2S)}$ denotes the world-average $\psi(2S)$ 
mass~\cite{pdg2006}.
For the $\psi(2S) \to J/\psi \pi^+ \pi^-$ candidates, 
$\Delta M \equiv M_{l^+ l^- \pi^+ \pi^-} - M_{l^+ l^-}$
is required to be within $0.580$ GeV/$c^2 < \Delta M  < 0.600$ GeV/$c^2$.
To reduce the fraction of incorrectly reconstructed 
$\psi(2S)$ signal candidates, 
we select $\pi^+ \pi^-$ pairs with 
an invariant mass greater than 400 MeV/$c^2$.
The $K^0_S$ selection criteria are the same as those described in 
Ref.~\cite{belle_b2s}; the invariant mass of the pion pairs is 
required to satisfy 0.482 GeV/$c^2 < M_{\pi^+\pi^-}  < 0.514$ GeV/$c^2$. 
%
%
\par We combine the $\psi(2S)$ and $K_S^0$ to form a neutral $B$
meson. The $B$ candidates are identified using 
two kinematic variables: the energy difference 
$\Delta E \equiv E_B^{\rm cms} - E_{\rm beam}^{\rm cms}$ and the
beam-energy-constrained mass 
$M_{\rm bc} \equiv \sqrt{(E_{\rm beam}^{\rm cms})^2 - (p_B^{\rm cms})^2}$,
where $E_{\rm beam}^{\rm cms}$ is the beam energy in the cms, and 
$E_B^{\rm cms}$ and $p_B^{\rm cms}$ are the cms energy and momentum, 
respectively, of the reconstructed $B$ candidate. 
In order to improve the $\Delta E$ resolution, the masses of the 
selected $J/\psi$ and $\psi(2S)$ candidates are constrained to their
nominal masses using mass-constrained kinematic fits.
For the $CP$ asymmetry fit, we select the candidates in the 
$\Delta E$-$M_{\rm bc}$ signal region defined as 
$|\Delta E| < 0.03$~GeV and
5.27 GeV/$c^2 < M_{\rm bc} <$ 5.29 GeV/$c^2$.
To suppress background from $e^+ e^- \to q \overline{q}$
($q = u, d, s,$~or~$c$) continuum events, we require that the 
event-shape variable $R_2$ be less than 0.5, where $R_2$ is the ratio of 
second to zeroth Fox-Wolfram moments~\cite{foxwolf}. 
%
%
\par The $b$ flavor of the accompanying $B$ meson is identified by a tagging 
algorithm~\cite{tag} that categorizes charged leptons, kaons, and $\Lambda$
baryons found in the event.
The algorithm returns two parameters: the 
$b$-flavor charge $q$, and $r$, which 
measures the tag quality
and varies from $r=0$ for no flavor discrimination to 
$r = 1$ for unambiguous flavor assignment.
If $r < 0.1$, the accompanying $B$ meson provides negligible
tagging information and we set the wrong tag probability to 0.5.
Events with $r > 0.1$ are divided into six $r$ intervals.
\par The vertex position for the $f_{CP}$ decay is reconstructed using charged
tracks that have a minimum number of SVD hits~\cite{tajima}. 
A constraint on the interaction point is also used 
with the selected tracks; the interaction point profile is convolved with the 
finite $B$-flight length in the plane perpendicular to the $z$ axis.
The pions from $K^0_S$ decays are not used for vertexing. 
The typical vertex reconstruction 
efficiency and $z$ resolution are 
95\% and 78~$\mu$m, respectively~\cite{belle_b2s}.
The $f_{\rm tag}$ vertex determination is obtained with 
well-reconstructed tracks that are not assigned to $f_{CP}$. 
The typical vertex reconstruction 
efficiency and $z$ resolution are 
93\% and 140~$\mu$m, respectively~\cite{belle_b2s}.
After all selection criteria are applied, we obtain 1618 and 1202 events 
for the $l^+ l^-$ and $J/\psi \pi^+ \pi^-$ modes in 
the $\Delta E$-$M_{\rm bc}$ fit region defined as 
5.2 GeV/$c^2 < M_{\rm bc} <$ 5.3 GeV/$c^2$ and $-0.1$
GeV $< \Delta E <$ 0.1 GeV, of which 680 and 712, respectively, are in the 
signal region.
%
%
%
\par
\begin{figure}[htbp]
\begin{center}
\begin{tabular}{c}
\includegraphics[width=9.4cm]{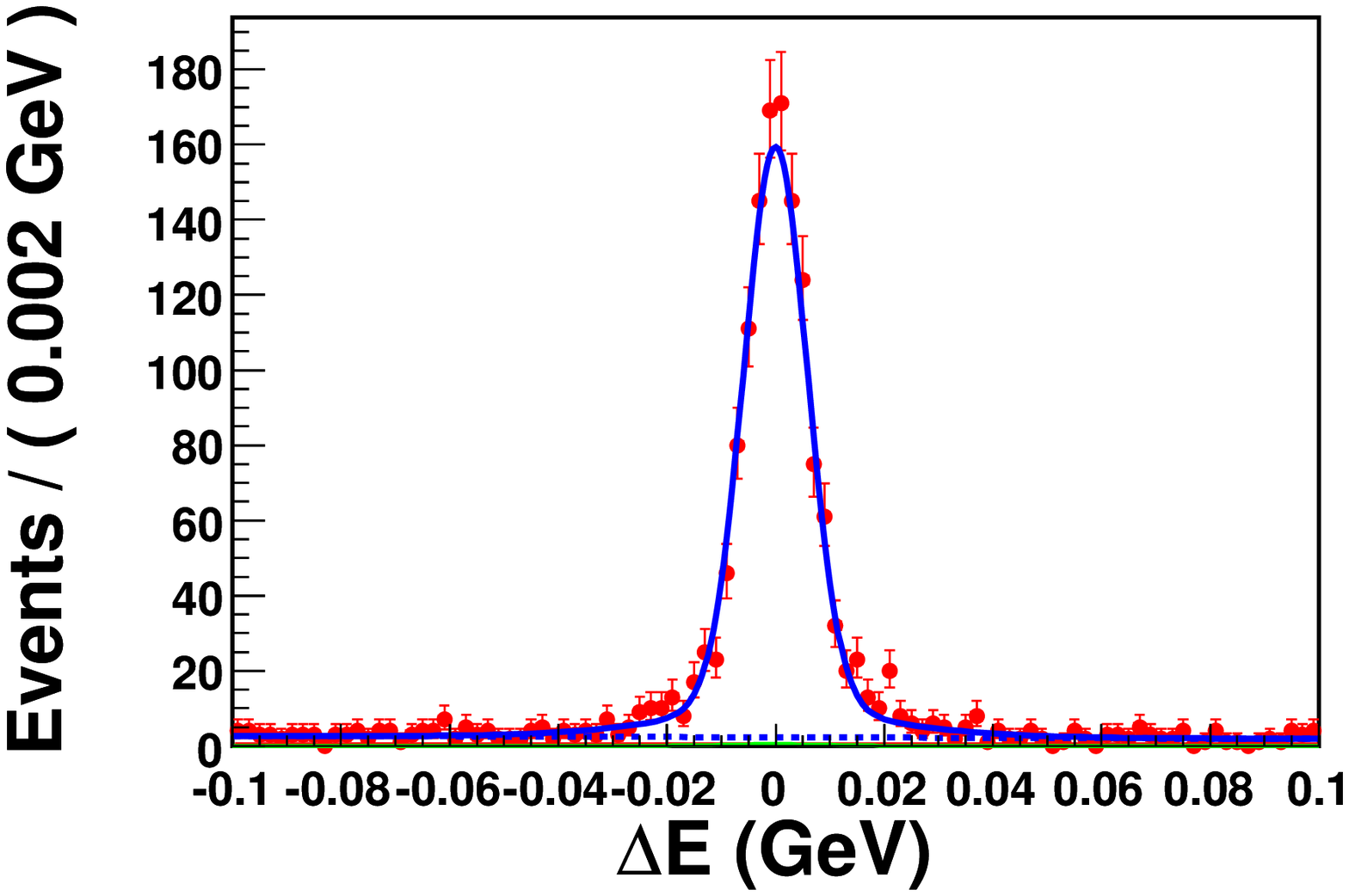} \\
\includegraphics[width=9.4cm]{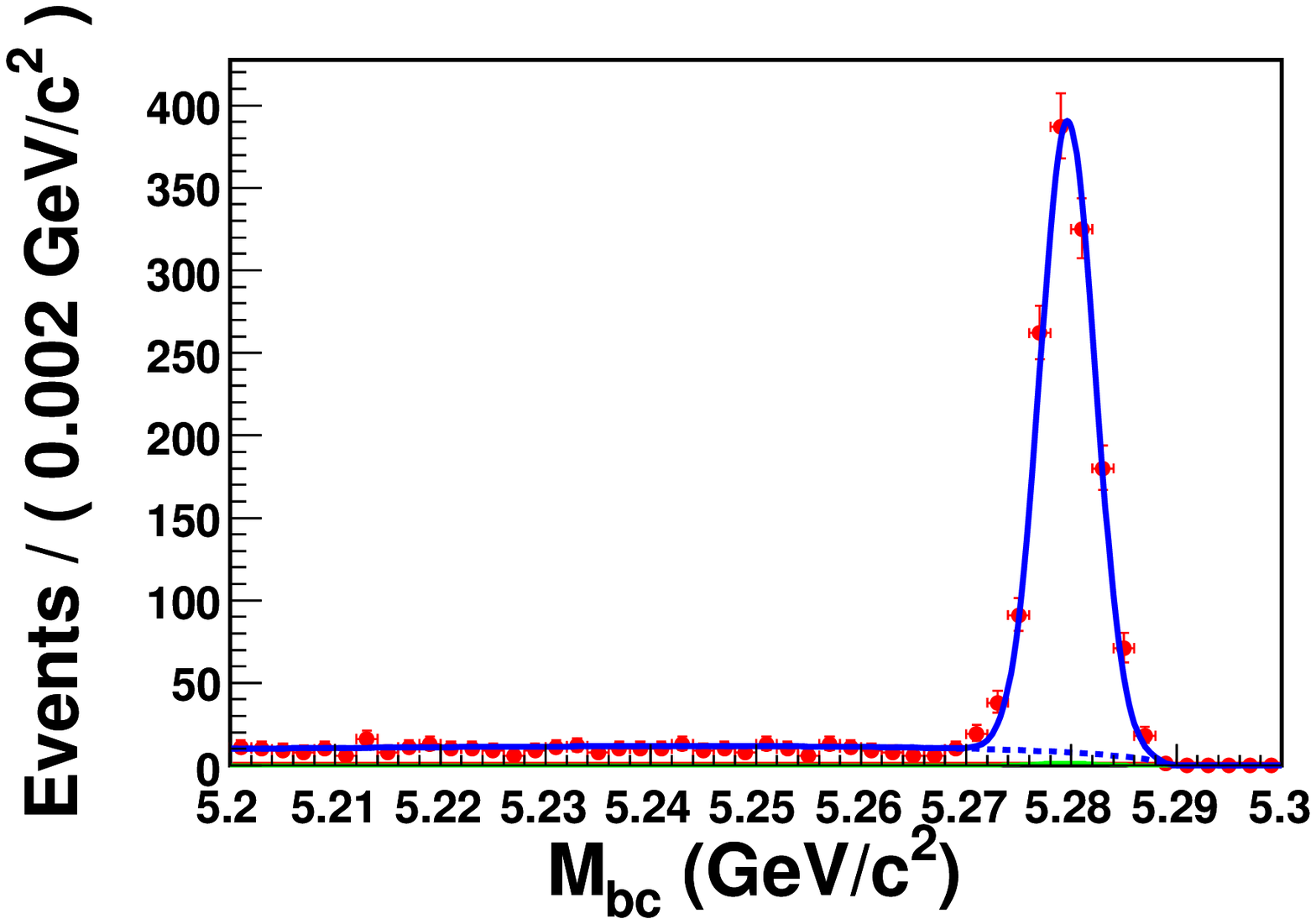} 
\end{tabular}
\end{center}
\caption{(color online). (a) $\Delta E$ distribution within the $M_{\rm bc}$ signal region, 
(b) $M_{\rm bc}$ distribution within the $\Delta E$ signal region for 
$B^0 \to \psi(2S) K_S^0$. The solid curves show the fits
to the signal plus background distributions, while the dashed curves show the
background contributions. The small contribution from peaking backgrounds
is discussed in the text.}
\rput[l]( -1.7, 13) {(a)}
\rput[l]( -1.7, 6.7) {(b)}
\label{fig_yield}
\end{figure}

 Figure~\ref{fig_yield} shows the reconstructed variables 
$\Delta E$ and $M_{\rm bc}$ after flavor tagging 
and vertex reconstruction. 
The signal yield is obtained from an extended
unbinned maximum-likelihood fit to the $\Delta E$-$M_{\rm bc}$ distribution. 
We model 
the shape for the signal component using the product of a 
double Gaussian for $\Delta E$ and a single Gaussian for $M_{\rm bc}$ 
whereas the combinatorial background is
described by the product of a first-order polynomial for $\Delta E$ and
an ARGUS~\cite{argus} function for $M_{\rm bc}$.
For the $\psi(2S) \to J/\psi \pi^+\pi^-$ mode, 
there is a background component that 
peaks like the signal (peaking background) in the $\Delta E$-$M_{\rm bc}$ 
signal region. This peaking background is mainly due to the
$J/\psi K_1(1270)^0$, $J/\psi K^*(892)^-\pi^+$ and 
$J/\psi K^0_S \pi^+ \pi^-$ modes, with no real 
$\psi(2S) \to J/\psi \pi^+ \pi^-$ in the final state. 
The fraction of such peaking events is 
estimated to be 1\% from the 
$\Delta M$ sidebands in data.
%
%
%
%
%
\begin{table}
\caption{Number of signal $B$ candidates ($\rm N_{sig}$)
and estimated signal purity ($p$) in the signal region
after flavor tagging and vertex reconstruction.}
\begin{ruledtabular}
\begin{tabular}{lcc}
Mode & $\;\; \rm N_{sig}\;\;$ & $\;p\;$ \\
\hline
$\psi(2S) (l^+l^-) K^0_S$              & $628 \pm 26$ & $0.92 \pm 0.01$ \\
$\psi(2S) (J/\psi \pi^+\pi^-) K^0_S$   & $656 \pm 26$ & $0.92 \pm 0.01$ \\
\end{tabular}
\end{ruledtabular}
\label{tab_yields}
\end{table}
%
%
%
The signal and background shapes for each decay mode are determined 
from Monte Carlo (MC) events;
these shapes are adjusted for small differences between MC and data
using a control sample of $B^+ \to \psi(2S) K^+$~\cite{conj} events, 
which have a final
state similar to the signal but with higher statistics. 
This sample, where no $CP$ asymmetry is expected, is also used to 
check the potential bias in the 
measurements of $CP$ violation parameters.
The signal yields and purity in the $\Delta E$-$M_{\rm bc}$ 
signal region after flavor tagging and vertex reconstruction 
are listed in Table~\ref{tab_yields}. We define the purity as 
the ratio of the signal yield to the total number of candidate events 
in the signal region.
%
%
%
\par We determine $\mathcal{S}_{f_{CP}}$ and 
$\mathcal{A}_{f_{CP}}$  by performing
an unbinned maximum-likelihood fit to the observed 
$\Delta t$ distribution for the candidate events in the signal region.
The likelihood function is
\begin{eqnarray}
\mathcal{L}(\mathcal{S}_{f_{CP}},\mathcal{A}_{f_{CP}})=
\prod_{i}\mathcal{P}_i(\mathcal{S}_{f_{CP}},\mathcal{A}_{f_{CP}};\Delta t_{i}),
\label{likelicpfiteq}
\end{eqnarray}
where the product 
includes events in the signal region.
We only use events with vertices that satisfy 
$|\Delta t| < 70$~ps
and $\xi<250$, where $\xi$ is the $\chi^2$ of the vertex fit 
calculated only in the $z$ direction.
The probability density function (PDF) is given by
\begin{eqnarray}
{\cal P}_i
&=& (1-f_{\rm ol}) \int 
\biggl[
f_{\rm sig} {\cal P}_{\rm sig}(\Delta t') R_{\rm sig}(\Delta t_i-\Delta t') 
\nonumber \\
&+&f_{\rm peak} {\cal P}_{\rm peak}(\Delta t') 
R_{\rm sig}(\Delta t_i-\Delta t')
\nonumber \\
&+&(1-f_{\rm sig}-f_{\rm peak}){\cal P}_{\rm bkg}(\Delta t')
R_{\rm bkg}(\Delta t_i-\Delta t')\biggr]
d(\Delta t')
\nonumber \\
&+&f_{\rm ol} P_{\rm ol}(\Delta t_i).
\label{eq:likelihood}
\end{eqnarray}
\par The signal fraction $f_{\rm sig}$ and the
peaking fraction $f_{\rm peak}$
depend on the $r$ region and are calculated on an 
event-by-event basis as a function of 
$\Delta E$ and $M_{\rm bc}$. The PDF for signal events, 
${\cal P}_{\rm sig}(\Delta t)$, is given by Eq.~\ref{eq_decay} 
and modified to incorporate the effect of incorrect flavor assignment; 
the parameters $\tau_{B^0}$ and $\Delta m_d$ 
are fixed to their world-average values~\cite{pdg2006}.
The distribution is then convolved with a resolution function 
$R_{\rm sig}(\Delta t)$ to take 
into account the finite vertex resolution.
The resolution function parameters, along with
the wrong tag fractions for the six $r$ intervals,
$w_l$ ($l = 1, 6$) and possible differences in $w_l$ 
between $B^0$ and $\overline{B}{}^0$
decays ($\Delta w_l$) are determined using a high-statistics control
sample of semileptonic and hadronic $b \to c$ 
decays~\cite{belle_cc,belle_b2s}. The PDF for non-peaking background events, 
${\cal P}_{\rm bkg}(\Delta t)$, is modeled as a sum of exponential 
and prompt components and is convolved with a sum of
two Gaussians, which parameterizes the resolution function 
$R_{\rm bkg}(\Delta t)$. Parameters in ${\cal P}_{\rm bkg}(\Delta t)$
and $R_{\rm bkg}(\Delta t)$ are determined from a fit 
to the $\Delta t$ distribution of events in the $\Delta E$-$M_{\rm bc}$ 
data sideband ($M_{\rm bc} < 5.26$ GeV/$c^2$, $-0.03$~GeV $< \Delta E <$ 
0.20~GeV). The PDF for peaking background events, 
${\cal P}_{\rm peak}(\Delta t)$,
is the same as ${\cal P}_{\rm sig}(\Delta t)$ with $CP$ parameters fixed
to zero. The term $P_{\rm ol}(\Delta t)$ is a broad Gaussian function that 
represents an outlier component with a small fraction $f_{\rm ol}$. 
The only free parameters in the final fit are ${\cal S}_{f_{CP}}$ and 
${\cal A}_{f_{CP}}$; these are determined by maximizing the likelihood 
function given by Eq.~\ref{likelicpfiteq}.
%
%
\par
\begin{figure}[htbp]
\begin{center}
\begin{tabular}{c}
\includegraphics[width=9.4cm]{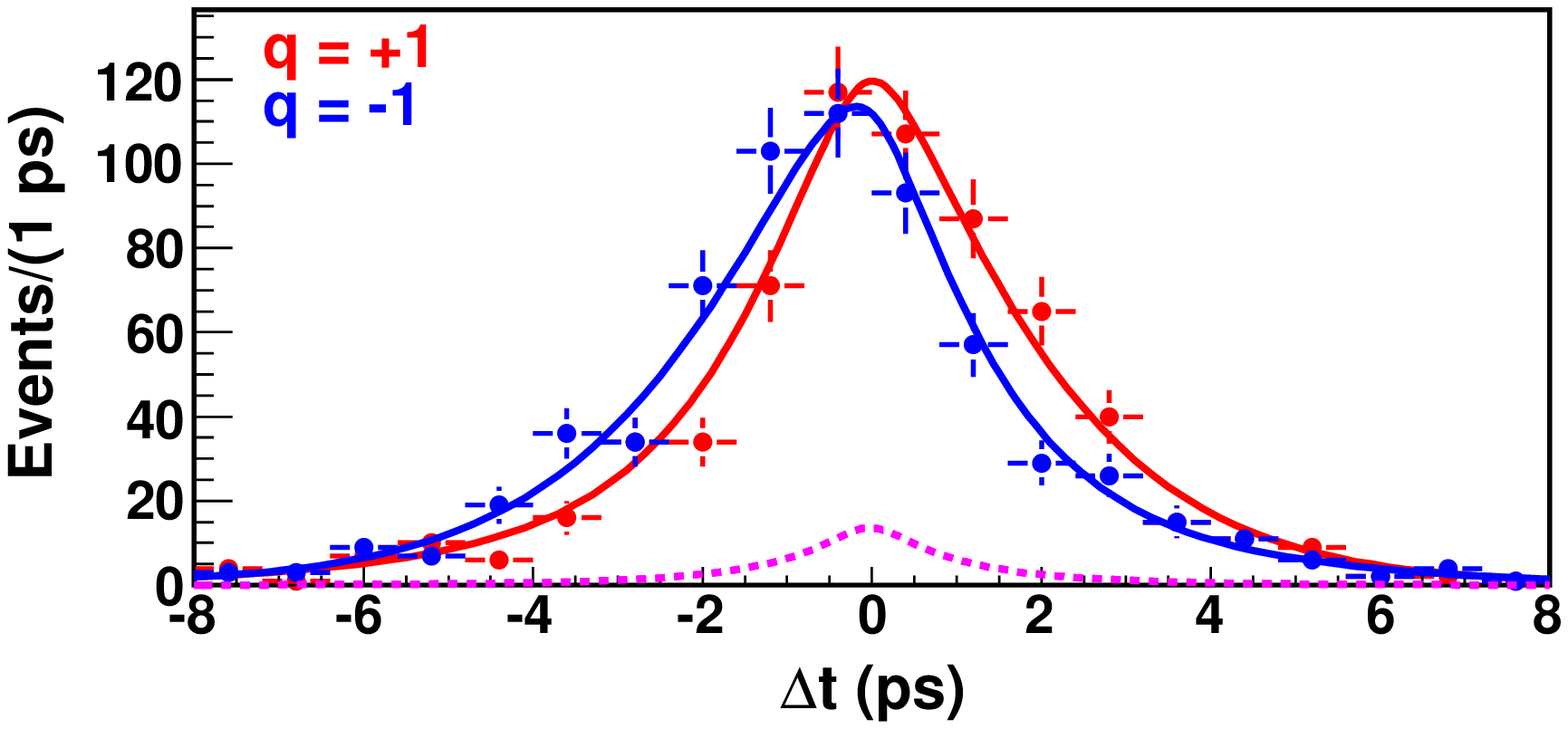} \\
\includegraphics[width=9.4cm]{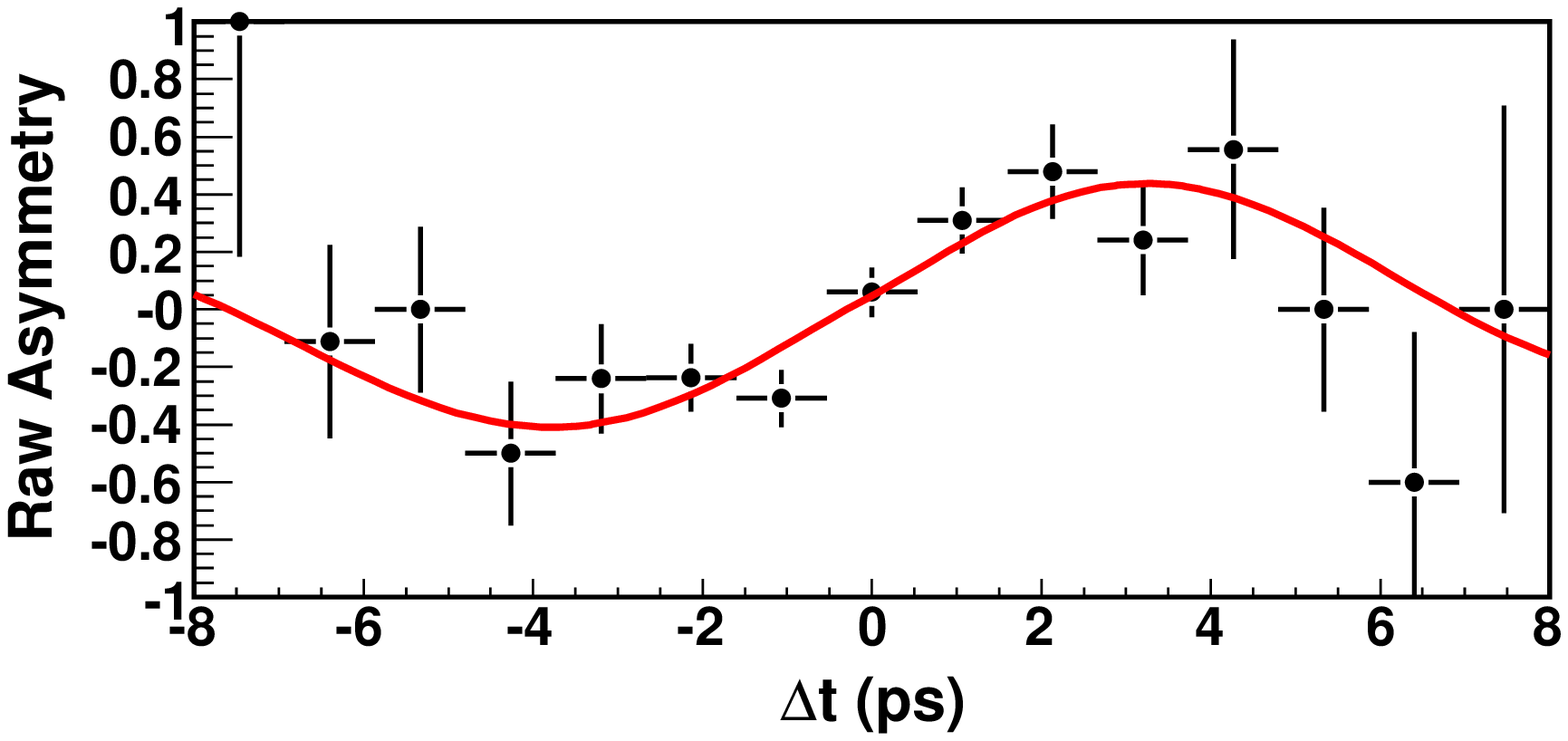} 
\end{tabular}
\end{center}
\caption{(color online). The top plot shows the $\Delta t$ distributions
for $q$ = $+1$ and $q$ = $-1$ 
with no requirement on $r$.
The dashed curve is the sum of 
backgrounds while the solid curves are the sum of signal and
backgrounds. The bottom plot is the raw asymmetry of 
well-tagged events ($r > 0.5$, 45\% of the total). 
The solid curve shows the result
of the unbinned maximum-likelihood fit. 
}
\label{raw_asym}
\end{figure}

 The unbinned maximum-likelihood fit to the
1300 events in the
signal region results in the $CP$ violation parameters,
\begin{eqnarray}
{\cal S}_{\psi(2S)K_S^0} = +0.72 \pm 0.09 (\rm stat) \pm 0.03 (\rm syst),
\nonumber \\
{\cal A}_{\psi(2S)K_S^0} = +0.04 \pm 0.07 (\rm stat) \pm 0.05 (\rm syst),
\nonumber
\end{eqnarray}
where the systematic uncertainties listed are described 
below.
We define the raw asymmetry in each $\Delta t$ bin by 
$(N_{+}-N_{-})/(N_{+}+N_{-})$, where $N_{+}$ $(N_{-})$
is the number of observed candidates with $q=+1$ $(-1)$.
Figure~\ref{raw_asym} shows the observed $\Delta t$ distributions 
for $q$ = $+1$ and $q$ = $-1$ 
with no requirement on the tagging quality (top), 
and the raw asymmetry 
for events with good tagging quality ($r > 0.5$) (bottom).
%
%

%
%
\par The systematic errors on ${\cal S}_{f_{CP}}$ and ${\cal A}_{f_{CP}}$,
summarized in Table~\ref{tab_syst}, are
evaluated by fitting the data
with each fixed parameter varied by its 1
standard deviation ($\sigma$) error.
The MC-determined parameters are varied by $\pm 2 \sigma$ 
to take into account
possible imperfect modeling in the MC.
We repeat the $CP$ fit procedure with 
the new value, add the differences in $S$, $A$ 
quadratically, and then assign the result as the systematic error.
The largest contribution 
to ${\cal S}_{f_{CP}}$ comes from vertex 
reconstruction (0.026). This includes the uncertainties in the 
interaction point profile (the smearing used to account for the $B$ flight
length is varied by $-10\;\mu{\rm m}$ and $+20\;\mu{\rm m}$), 
the tag side track selection
criteria, the helix parameter correction, the $|\Delta t|$ range 
(varied by $\pm$30 ps), the vertex quality cut $\xi$ 
(changed to $\xi<150$ and $\xi<500$), the $\Delta z$ measurement, and 
imperfect SVD alignment. The last two are obtained from 
a study of $J/\psi K_S^0$.
Each physics parameter ($\tau_{B^0}$, $\Delta m_d$) is varied
by the error in its world-average value~\cite{pdg2006}.
Systematic errors due to uncertainties in wrong tag fractions are estimated
by varying the parameters $w_l$, $\Delta w_l$ in each $r$ region
by their $\pm 1 \sigma$ errors. Systematic
errors due to uncertainties in the resolution function are 
estimated by varying 
each resolution parameter obtained from data (MC) by 
$\pm 1 \sigma$ ($\pm 2 \sigma$). The $\Delta E$, $M_{\rm bc}$
parameters and signal fraction in each $r$ region
are varied to estimate the systematic errors. No significant bias is seen 
by fitting a large sample of MC events. The systematic errors from 
uncertainties in the 
peaking background are obtained by varying the 
peaking fraction, shape, as well as its $CP$ asymmetry parameters. 
The systematic errors from uncertainties 
in the background
$\Delta t$ shape are estimated by varying each background parameter 
by its statistical error. We also include the effects of interference
between CKM-favored and CKM-suppressed $B \to D$ transitions in the
$f_{\rm tag}$ final state~\cite{long}. We add each contribution 
above in quadrature to obtain the total systematic uncertainty.
%
%
%
\par We perform various cross-checks for this measurement.
A fit to the first data sample (SVD-I) results in the 
CP violation parameter, ${\cal S} = 0.97 \pm 0.18$,
which is consistent with our previous result~\cite{belle_cc}.
A fit to the $CP$ asymmetries of the control sample gives the 
$CP$ violation parameters, 
${\cal S} = 0.02 \pm 0.05$ and 
${\cal A} = -0.03 \pm 0.03$,
which are consistent with no $CP$ asymmetry.
A fit to the sideband events in the
$B^0 \to \psi(2S) K_S^0$ data sample 
gives an asymmetry consistent with zero 
(${\cal S} = 0.02 \pm 0.21$, ${\cal A} = -0.04 \pm 0.10$).
A lifetime fit to $B^0 \to \psi(2S) K_S^0$ and $B^+ \to \psi(2S) K^+$
gives $\tau_{B^0} = 1.51 \pm 0.05$ ps and 
$\tau_{B^+} = 1.62 \pm 0.03$ ps, respectively, which are consistent 
with the world-average values.
We also examine the $CP$ violation parameters separately for the 
$\psi(2S) \to l^+l^-$ 
(${\cal S} = 0.84 \pm 0.13$, ${\cal A} = 0.14 \pm 0.09$)
and $\psi(2S) \to J/\psi \pi^+\pi^-$ 
(${\cal S} = 0.61 \pm 0.13$, ${\cal A} = -0.09 \pm 0.10$)
decay modes. We find that 
all results are consistent within errors.
%
%
%
%
\begin{table}
\caption{Systematic uncertainties.}
\begin{center}
\begin{ruledtabular}
\begin{tabular}{lcc}
Parameter  & $\;\Delta {\cal S}_{\psi(2S) K_S^0}\;$ & 
$\;\Delta {\cal A}_{\psi(2S) K_S^0}\;$\\
\hline
Vertexing                   & 0.026 & 0.028 \\
Wrong tag fraction          & 0.006 & 0.023 \\
Resolution function         & 0.007 & 0.005 \\
Fit bias                    & 0.012 & 0.011  \\
Physics parameters          & 0.001 & 0.001 \\
Peaking background          & 0.006 & 0.005 \\
PDF shape and fraction      & 0.001 & 0.003 \\
Background $\Delta t$ shape & 0.003 & 0.003 \\
Tag side interference       & 0.001 & 0.036 \\
\\
Total                       & 0.031 & 0.053 \\
\end{tabular}
\end{ruledtabular}
\end{center}
\label{tab_syst}
\end{table}
%
%
%
\par In summary, we have performed improved measurements of $CP$ violation
parameters $\sin 2 \phi_1$ and ${\cal A}_{f_{CP}}$ for 
$B^0 \to \psi(2S) K_S^0$ using $657 \times 10^6$ $B\overline{B}$ events.
These measurements supersede our previous result~\cite{belle_cc}  
and are in agreement 
with results from measurements of 
$B^0 \to J/\psi K^0$~\cite{hfag}.
Combining the results from $B^0 \to J/\psi K^0$~\cite{belle_new}
and $B^0 \to \psi(2S) K_S^0$ decays, we obtain a new Belle
average $\sin 2 \phi_1 = 0.650 \pm 0.029 \pm 0.018$.
\par We thank the KEKB group for excellent operation of the
accelerator, the KEK cryogenics group for efficient solenoid
operations, and the KEK computer group and
the NII for valuable computing and Super-SINET network
support.  We acknowledge support from MEXT and JSPS (Japan);
ARC and DEST (Australia); NSFC (China); 
DST (India); MOEHRD, KOSEF and KRF (Korea); 
KBN (Poland); MES and RFAAE (Russia); ARRS (Slovenia); SNSF (Switzerland); 
NSC and MOE (Taiwan); and DOE (USA).


\begin{thebibliography}{99}
\bibitem{ckm} 
N.~Cabibbo, Phys. Rev. Lett. {\bf 10}, 531 (1963);
M.~Kobayashi and T.~Maskawa, Prog. Theor. Phys. {\bf 49}, 652 (1973).
\bibitem{carter} 
A.~B.~Carter and A.~I.~Sanda, Phys. Rev. D {\bf 23}, 1567 (1981);
I.~I.~Bigi and A.~I.~Sanda, Nucl. Phys. B {\bf 193}, 85 (1981).
\bibitem{beta} Another naming convention 
$\beta$(= $\phi_1$) is also used in the literature.
\bibitem{belle_cc}  
K.~Abe {\it et al.} (Belle Collaboration), 
Phys. Rev. D {\bf 71}, 072003 (2005).
\bibitem{belle_new}  
K-F.~Chen {\it et al.} (Belle Collaboration), 
Phys. Rev. Lett. {\bf 98}, 031802 (2007).
\bibitem{babar_cc} 
B.~Aubert {\it et al.} (BaBar Collaboration), Phys.
Rev. Lett. {\bf 99}, 171803 (2007).
\bibitem{impro}This analysis is improved compared to our
previous publication~\cite{belle_cc}. 
The peaking backgrounds are estimated directly from the data sidebands. 
The signal and peaking fraction in the time-dependent PDF 
are determined for each $r$-bin. In addition to the lepton 
tracks from the $J/\psi$, we use the prompt $\pi^+\pi^-$ 
tracks in the vertex reconstruction for the
$\psi(2S)(J/\psi \pi^+ \pi^-)K_S^0$ decay mode. 
Only lepton tracks from the $J/\psi$ were used 
in the previous analysis.
\bibitem{kekb}
S.~Kurokawa and E.~Kikutani, 
Nucl. Instrum. Methods Phys. Res., Sect. A 
{\bf 499}, 1 (2003),
and other papers included in this volume.
\bibitem{Belle}
A.~Abashian {\it et al.} (Belle Collaboration),
Nucl. Instrum. Methods Phys. Res., Sect. A 
{\bf 479}, 117 (2002).
\bibitem{Natkaniec} Z.~Natkaniec (Belle SVD2 Group),
Nucl. Instrum. Methods Phys. Res., Sect. A 
{\bf 560}, 1 (2006).
\bibitem{pdg2006} W.-M.~Yao {\it et al.}, J. Phys. G {\bf 33}, 1 (2006).
\bibitem{belle_b2s}
K-F.~Chen {\it et al.} (Belle Collaboration), Phys. Rev. D {\bf 72},
012004 (2005).
\bibitem{foxwolf} G.~C.~Fox and S.~Wolfram, Phys.
Rev. Lett. {\bf 41}, 1581 (1978).
\bibitem{tag}
H.~Kakuno {\it et al.}, 
Nucl. Instrum. Methods Phys. Res., Sect. A 
{\bf 533}, 516 (2004).
\bibitem{tajima}
H.~Tajima {\it et al.}, Nucl. Instrum. Methods Phys. Res., 
Sect. A {\bf 533}, 370 (2004).
\bibitem{argus} H.~Albrecht {\it et al.} (ARGUS Collaboration), Phys. 
Lett. B {\bf 241}, 278 (1990).
\bibitem{conj} Throughout this paper, the inclusion of the charge-conjugate 
decay mode is implied unless otherwise stated.
\bibitem{long}
O.~Long, M.~Baak, R.~N.~Cahn and D.~Kirkby, Phys. Rev. D {\bf 68}, 
034010 (2003).
\bibitem{hfag} 
See http://www.slac.stanford.edu/xorg/hfag/ for updated results.

\end{thebibliography}
\end{document}